\documentclass[english]{iopart}
\usepackage[T1]{fontenc}
\usepackage[latin9]{inputenc}
\usepackage{graphicx}

\newcommand{\lyxdot}{.}

\usepackage{iopams}
\usepackage{setstack}

\usepackage{babel}

\begin{document}

\title{Ninja data analysis with a detection pipeline based on the Hilbert-Huang
Transform }

\author{Alexander Stroeer$^{1,2}$, Jordan Camp$^{1}$ }

\address{$^{1}$NASA/GSFC, Code 663, Greenbelt, MD 20771; $^{2}$CRESST, University
of Maryland, College Park, MD 20742}

\ead{Alexander.Stroeer@nasa.gov}
\begin{abstract}
The Ninja data analysis challenge allowed the study of the sensitivity
of data analysis pipelines to binary black hole numerical relativity
waveforms in simulated Gaussian noise at the design level of the LIGO
observatory and the VIRGO observatory. We analyzed NINJA data with
a pipeline based on the Hilbert Huang Transform, utilizing a detection
stage and a characterization stage: detection is performed by triggering
on excess instantaneous power, characterization is performed by displaying
the kernel density enhanced (KD) time-frequency trace of the signal.
Using the simulated data based on the two LIGO detectors, we were
able to detect 77 signals out of 126 above SNR 5 in coincidence, with
43 missed events characterized by signal to noise ratio SNR<10. Characterization
of the detected signals revealed the merger part of the waveform in
high time and frequency resolution, free from time-frequency uncertainty.
We estimated the timelag of the signals between the detectors based
on the optimal overlap of the individual KD time-frequency maps, yielding
estimates accurate within a fraction of a millisecond for half of
the events. A coherent addition of the data sets according to the
estimated timelag eventually was used in a final characterization
of the event. 
\end{abstract}
\maketitle

\section{Introduction}

The Numerical INJection Analysis project (NINJA,\cite{ninja}) allowed
the study of the sensitivity of data analysis (DA) pipelines to binary
black hole numerical relativity (NR) waveforms in simulated Gaussian
noise at the design level of the LIGO observatory \cite{barish1999lad,sigg2006sld}
and the VIRGO observatory \cite{bradaschia1990vpw,caron1997vig}.
The project combined for the first time numerical relativity simulations
with gravitational wave data analysis strategies to create a realistic
sensitivity study. Overall, NINJA saw 65 participants from 23 institutions,
with 10 NR groups contributing waveforms of their choice, and 9 DA
teams analyzing the data with various methods that included modelled
approaches (like matched filtering or Bayesian strategies like Markov
Chain Monte Carlo techniques or Bayesian model estimators), and unmodelled
approaches (like the Q-transform which utilizes sine-gaussians with
varying number of oscillations as basis set of a transformation)\cite{ninja}.
The original numerical results for the NINJA numerical waveform contributions
are described in \cite{Hannam:2007ik,Hannam:2007wf,Tichy:2008du,Marronetti:2007wz,Tichy:2007hk,Pollney:2007ss,Rezzolla:2007xa,Vaishnav:2007nm,Hinder:2007qu,Buonanno:2006ui,Pretorius:2007jn,Boyle:2007ft,Scheel:2008rj,Etienne:2007hr}
(where these are published results), the codes are described in \cite{Brugmann:2008zz,Husa:2007hp,Koppitz:2007ev,Pollney:2007ss,Imbiriba:2004tp,vanMeter:2006vi,Zlochower:2005bj,Campanelli:2005dd,Sperhake:2006cy,Hinder:2007qu,Pretorius:2004jg,Pretorius:2005gq,Scheel:2006gg,Etienne:2007hr}.

The Goddard LSC group applied an unmodelled pipeline based on the
Hilbert-Huang Transform (HHT) \cite{huang2005hht,camp2007ahh} to
the analysis of NINJA data. Because our pipeline \cite{hhtatlowsnr2008}
was very recently developed, and is still being tested, we chose to
concentrate solely on the analysis of the LIGO Hanford and Livingston
data sets. The HHT is an adaptive algorithm that decomposes the data
into Intrinsic Mode Functions (IMF's), each representing a unique
locally monochromatic frequency scale of the data, with the original
data recovered if summed over all IMF's. The Hilbert transform as
applied to each IMF reveals the instantaneous frequency (IF) and instantaneous
power (IP) as a function of time, providing high time-frequency resolution
to detected signals without the usual time-frequency-uncertainty as
found in basis set methods like the Fourier transform.

\section{Methods}

\begin{figure}
\includegraphics[scale=0.5]{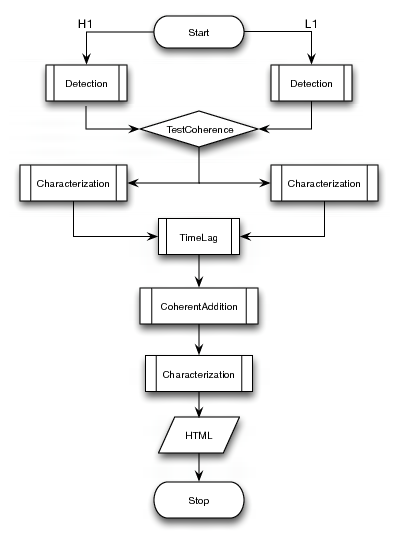}\includegraphics[scale=0.5]{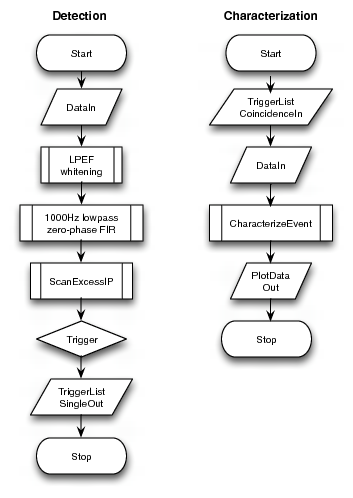}

\caption{Flowchart for the HHT data analysis pipeline. }

\label{Flo:HHTflowchart} 
\end{figure}

We applied an automatic two-stage HHT pipeline to detect and characterize
a signal as follows (see flowchart in Fig. \ref{Flo:HHTflowchart}).
The data was pre-processed with the use of a whitening linear predictor
error filter followed by a 1000 Hz low pass zero-phase Finite Impulse
Response (FIR) filter. In the detection stage (here within subroutine
{}``ScanExcessIP''), the IP's from each detector are divided into
blocks with similar statistical properties according to the Bayesian
Block algorithm \cite{mcnabb2004obe}. The presence of excess power
in a block generates a trigger and the search for a detector coincidence,
with triggered blocks yielding detection statistics, start and end
times, the maximum signal frequency, and the signal- to-noise ratio
(SNR) of the signal. The characterization stage (here within subroutine
{}``CharacterizeEvent'') uses information from the detection stage
to filter the event, and then zooms into the signal region of interest
and derives the 1) IF, 2) highly detailed time-frequency-power (tfp)
maps and 3) weighted (with power) kernel density time-frequency maps
(KD time-frequency map). The detailed approach behind both these subroutines
can be found in \cite{hhtatlowsnr2008}, and are not further discussed
in this paper (see also \cite{camp2007ahh} for further reference).

Our pipeline analyzes each detector separately over a time window
of 1024 points, corresponding to 250 msec at the NINJA data sampling
rate. A coincidence test on the individual detector triggers is performed
by a simple time window analysis at first, set over the full 250 msec
window to account for uncertainties in the timing of the event. If
the separate detector triggers are within this window, the overlap
of the individual time-frequency's is finally used to determine coincidence
and to estimate the timelag between the signals. The estimation of
timelags is performed by shifting one KD time-frequency map with respect
to the other in increments of the sampling time of NINJA ($1/4096$Hz)
, multiplying the maps and summing over the multiplied values. The
maximum of the resulting distribution is the estimated timelag; if
this timelag is within +/-10 msec (the light travel time between Hanford
and Livingston) a coincidence is established. Finally we construct
a coherent addition of the two detector data streams used in a final
characterization of the signal.

\section{Results}

\begin{figure}
\includegraphics[scale=0.5]{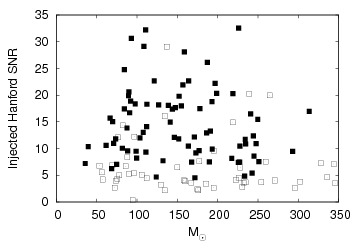}\includegraphics[scale=0.5]{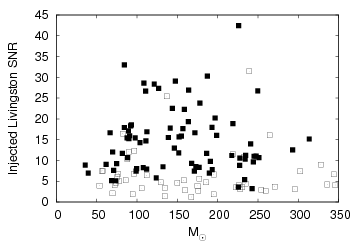}

\caption{Found (filled blocks) and missed (empty diamonds) events in the total
mass (solar units)/SNR plane per detector}

\label{Flo:mc.vs.snr} 
\end{figure}

We were able to identify 77 out of 126 events in coincidence between
Hanford and Livingston. Out of the 49 missed, 38 are SNR<10, 5 are
SNR<10 in one detector and SNR>10 in the other, 6 were SNR>10 (see
Fig. \ref{Flo:mc.vs.snr}). We therefore reason that most of the of
missed events are low SNR cases in which no blocks were triggered.
The 6 missed events at high SNR were caused by a timing error in the
coincidence test, and are not subject to the specifics of the injected
waveform. The pipeline detection threshold setting allowed 3 noise
coincidences over the $10^{5}$ sec data set, or a false alarm rate
of $\sim10^{-2}$ Hz for each detector. We did not attempt to use
vetoes in our analysis.

Overall, as we show in single examples below, the triggered blocks
frame the visible signal accurately, therefore providing a high sensitivity
detection stage as power from noise is not included in the detection
statistic which basically sums over the triggered blocks. However,
we found the derived SNR over the data range of the triggered blocks
to be generally underestimated, indicating that the HHT sees mainly
the peaked merger part of the waveform, and tends to be less efficient
at capturing the lower amplitude inspiral and the ringdown of the
signal which would normally contribute to the SNR estimation. This
leads us to employ the initial 250 msec coincidence window mentioned
above before tightening the coincidence condition with the KD time-frequency
maps.

\begin{figure}
\includegraphics[scale=0.5]{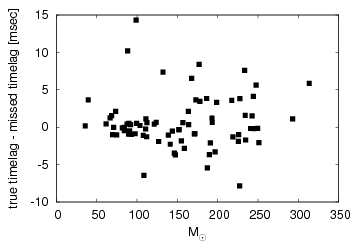}

\caption{The timelag estimate as displayed against the total mass (solar units)}

\label{Flo:timelagMchirp} 
\end{figure}

\begin{figure}
\includegraphics[scale=0.5]{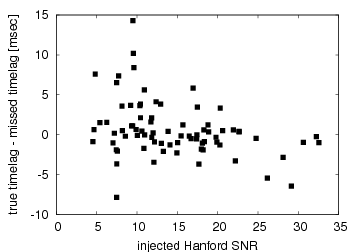}\includegraphics[scale=0.5]{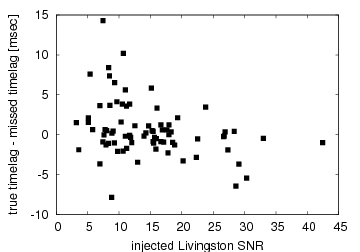}

\caption{The timelag estimate as displayed against the SNR per detector}

\label{Flo:timelagSNR} 
\end{figure}

The differences between true timelags and the detected timelag between
the detectors are plotted versus total mass (solar units) and individual
SNR in Fig. \ref{Flo:timelagMchirp} and Fig. \ref{Flo:timelagSNR}
respectively. The timelags of 18 events were estimated with an accuracy
less than 0.5msec, 14 events with an accuracy of order 1 msec, 16
of order 2 msec, 6 of order 3 msec and 23 larger than 4 msec. We found
evidence that uncertainties in timelag estimates are smaller for large
SNR (>10) in both detectors and also for smaller total mass (solar
units), which yields a shorter, more peaked waveform favorable to
the detection strategy of the HHT pipeline.

\begin{figure}
\includegraphics[scale=0.285]{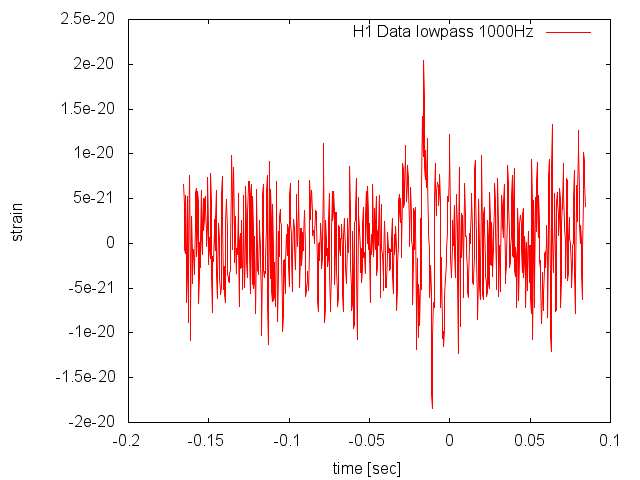}\includegraphics[scale=0.285]{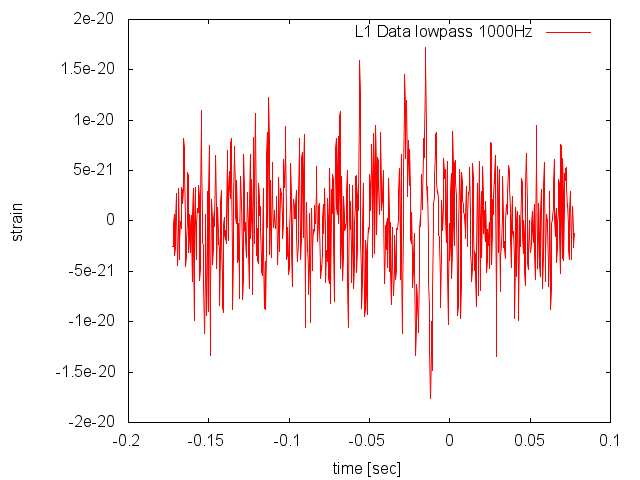}

\caption{The data of event 894376744, in Hanford (left panel) and Livingston
(right panel).}

\label{Flo:Data894376744} 
\end{figure}

\begin{figure}
\includegraphics[scale=0.285]{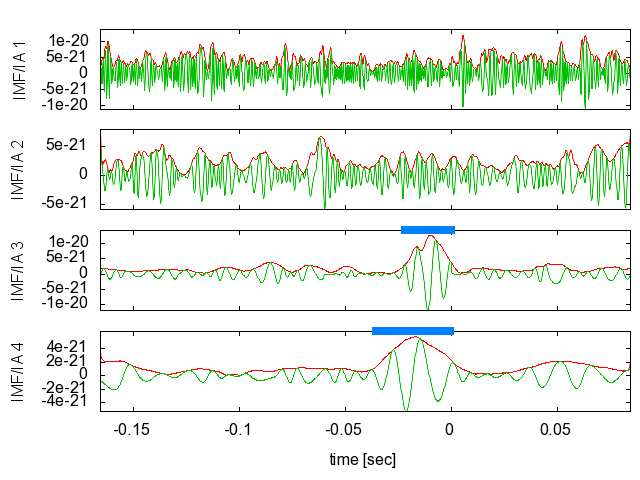}\includegraphics[scale=0.285]{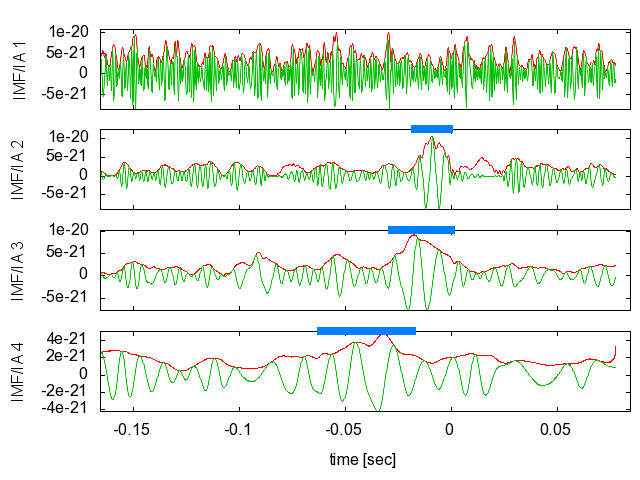}

\caption{The EMD decomposition (red) and the IP (green) of event 894376744,
in Hanford (left panel) and Livingston (right panel). Triggered blocks
are seen in blue.}

\label{Flo:Data894376744-imf} 
\end{figure}

\begin{figure}
\includegraphics[scale=0.285]{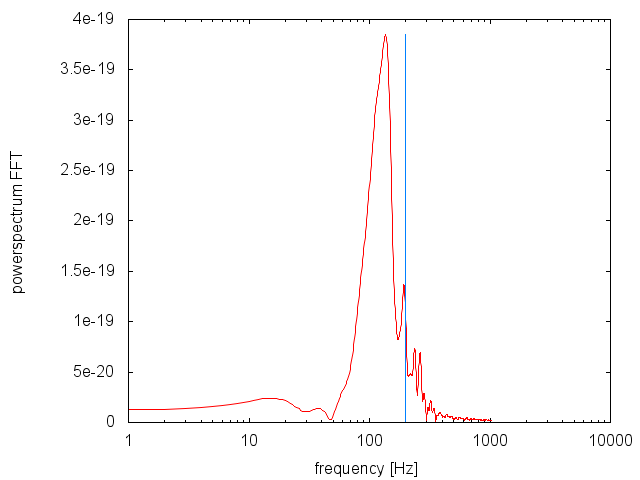}\includegraphics[scale=0.285]{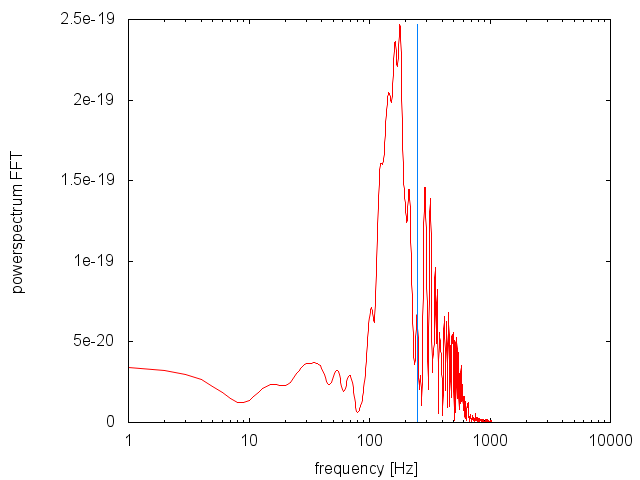}

\caption{The estimation of the maximum frequency of event 894376744, in Hanford
(left panel) and Livingston (right panel). The estimated upper frequency
is seen as blue line.}

\label{Flo:Data894376744-freqmax} 
\end{figure}

\begin{figure}
\includegraphics[scale=0.285]{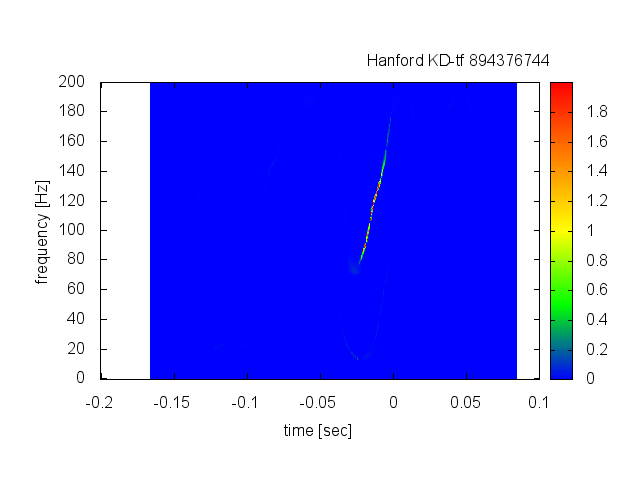}\includegraphics[scale=0.285]{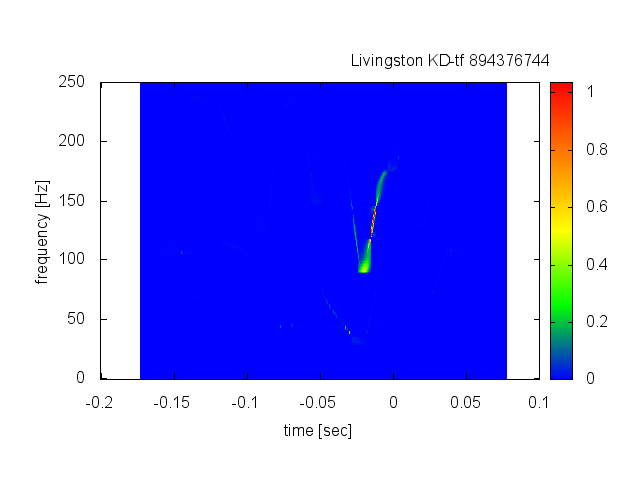}

\caption{The KD time-frequency map (time-frequency) of event 894376744, in
Hanford (left panel) and Livingston (right panel). }

\label{Flo:Data894376744-wkde.w.e} 
\end{figure}

\begin{figure}
\includegraphics[scale=0.285]{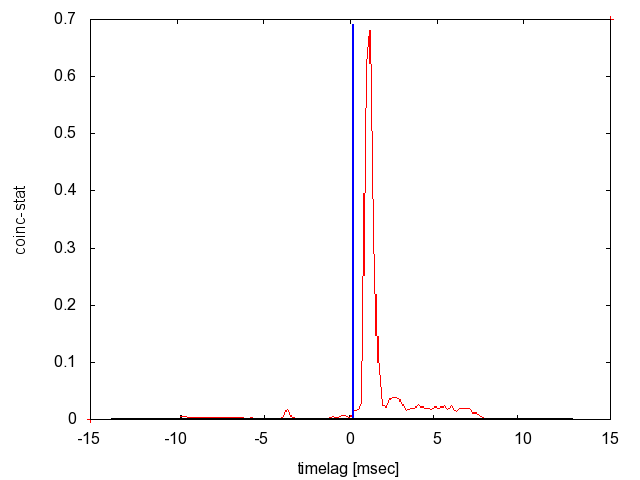}\includegraphics[scale=0.285]{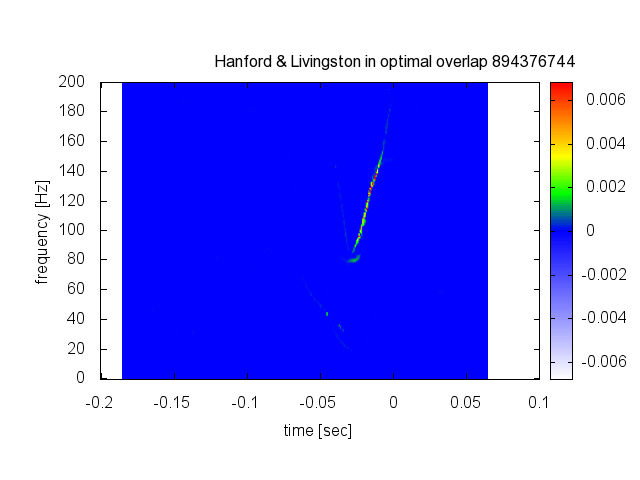}

\caption{Shifting the individual Hanford and Livingston KD time-frequency maps
of event 894376744 by increments of the NINJA sampling time to a total
of +/- 10 msec, multiplying the overlapping maps and finally summing
over the individual multiplied values, we define a statistic that
measures the quality of the overlap. The optimal overlap of the KD
time-frequency maps of Hanford and Livingston of event 894376744 is
the timelag corresponding to the peak (left panel). The real timelag
is shown in blue. The right panel shows the corresponding overlap
in the KD-tf map.}

\label{Flo:Data894376744-timelag} 
\end{figure}

We turn now to a discussion of single events to display the individual
stages of our pipeline, and to illustrate the advantages but also
some unresolved issues with our approach. First we concentrate on
a medium SNR event, event 894376744, with explicit SNR in Hanford
of 9.348 and SNR in Livingston of 14.72. Fig. \ref{Flo:Data894376744}
shows the time-series of the signal in noise. The EMD decomposition
(in red) and the IP (green) of the event is visible in Fig. \ref{Flo:Data894376744-imf},
with triggered blocks indicated in blue above the individual panels.
We find in this plot a demonstration of the use of triggered blocks
to accurately frame regions of excess power. The estimation of the
maximum frequency of the event utilizes triggered blocks to select
the times over which a Fast Fourier Transform and a power spectrum
estimate is performed, by taking the lowest IMF which was triggered
and selecting the region within the triggered block (see Fig. \ref{Flo:Data894376744-freqmax}).
The maximum frequency is estimated by locating the transition of the
power spectrum from signal power to noise power (found by noting the
first inflection point of the power spectrum after the maximum of
the signal spectrum and adding a small increment in frequency, of
order 50 Hz, for conservatism). Since the block region is very short,
of order tens of msec, we experience time-frequency uncertainties
widening the power spectrum, leading to a related bias in the estimate
of the upper frequency. Using the derived maximum frequency of the
signal, we aggressively filter (low-pass) the data so that we can
obtain accurate IFs and KD time-frequency maps. The KD time-frequency
map of the event in Hanford and Livingston is seen in Fig. \ref{Flo:Data894376744-wkde.w.e}.
The precision of the overlap analysis of the individual time-frequency
is apparent; the estimated timelag error, subject to the uncertainty
of the individual KD time-frequency maps, is only 1 msec (see Fig.
\ref{Flo:Data894376744-timelag}). An objective analysis of the accuracy
of the derived time-frequency evolution cannot be given in this proceeding
as waveforms and detailed time frequency evolutions of the injected
waveform were not yet released by the individual NR groups; they will
be given at a different location (see \cite{ninja}).

\begin{figure}
\includegraphics[scale=0.285]{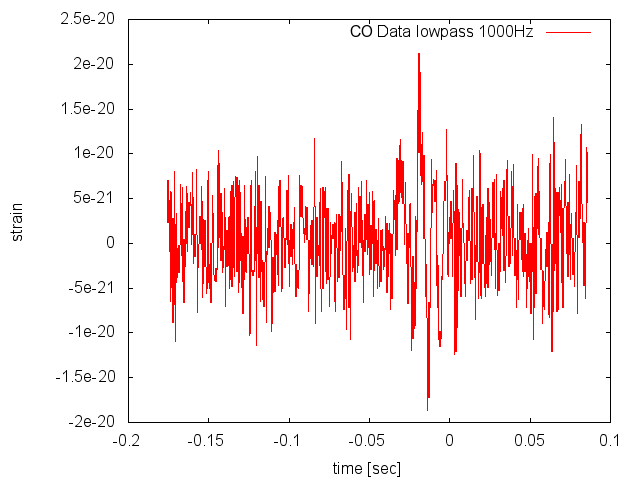}\includegraphics[scale=0.285]{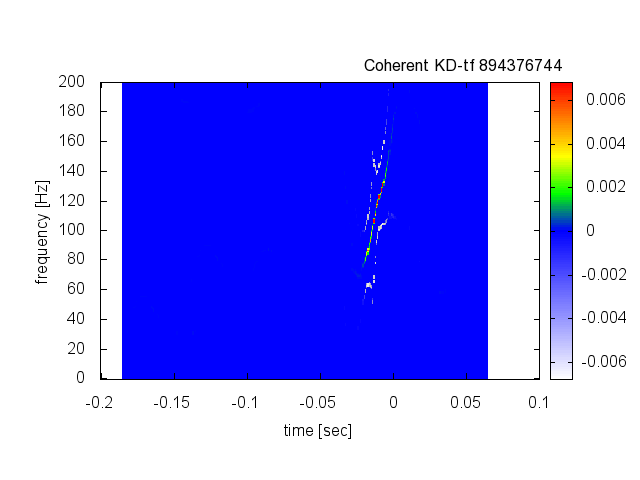}

\caption{The coherent addition of the data according to estimated timelag of
event 894376744 is seen. }

\label{Flo:Data894376744-coherent} 
\end{figure}

With the timelag between the signals determined, a coherent addition
of the signals can be made. Since the accuracy of the HHT decomposition
depends strongly on the signal amplitude, a coherent addition can
significantly improve the signal characterization. Fig. \ref{Flo:Data894376744-coherent}
shows the coherent addition of the data sets and its final characterization,
here shown with uncertainty estimates (for details , see \cite{hhtatlowsnr2008}).
The most significant source of error in the coherent analyses is the
uncertainty in the timelag estimate, and this coupling remains under
study.

\begin{figure}
\includegraphics[scale=0.285]{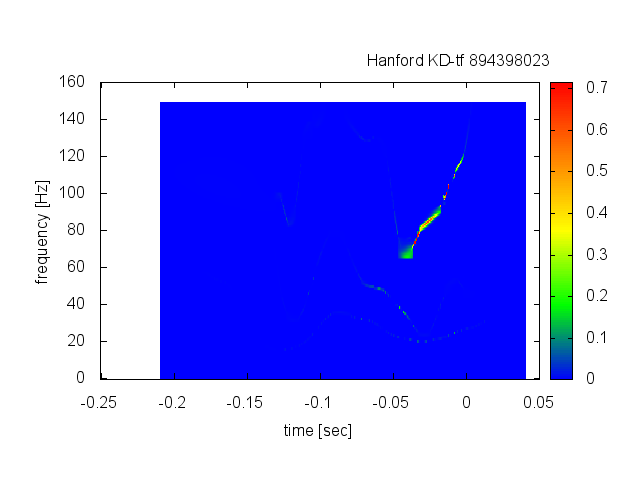}\includegraphics[scale=0.285]{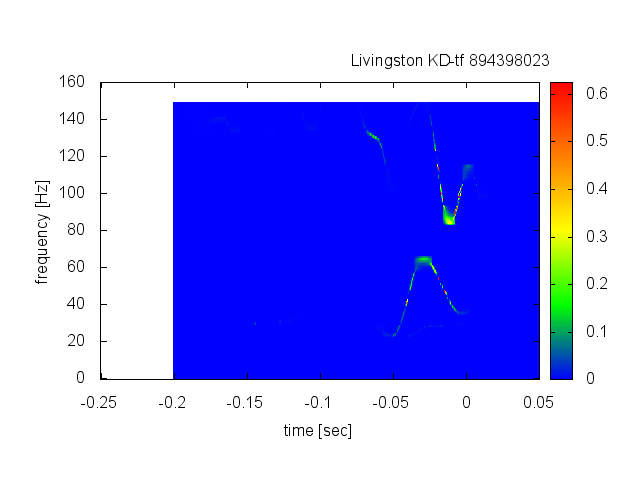}

\caption{The KD time-frequency map (time-frequency) of event 894398023, in
Hanford (left panel) and Livingston (right panel). }

\label{Flo:Data894398023-1} 
\end{figure}

\begin{figure}
\includegraphics[scale=0.285]{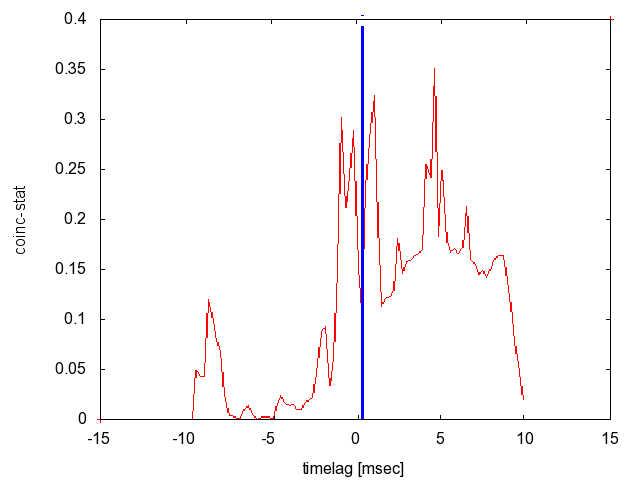}\includegraphics[scale=0.285]{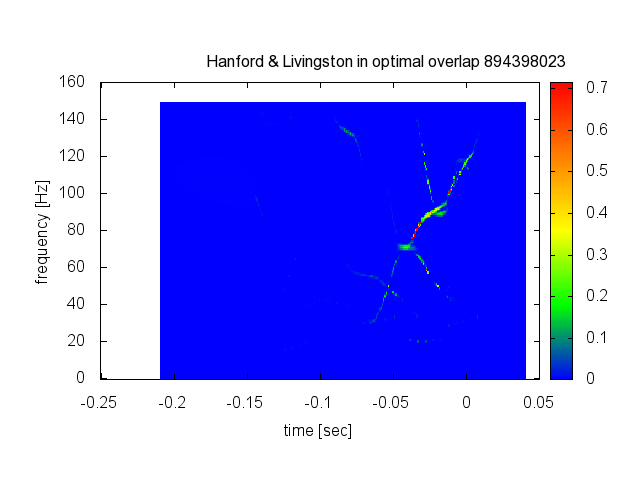}

\caption{The timelag estimate of event 894398023 shows uncertainties caused
by the occurrence of mode mixing in one of the detectors (left panel).
As it becomes visible in the right panel, this is caused by the inability
to perfectly overlap the two traces in the KD-tf map.}

\label{Flo:Data89439802timelag} 
\end{figure}

The precise estimation of timelags requires an accurate measure of
the time-frequency evolution in both detectors, so that there is minimal
uncertainty in the time-frequency overlap. Thus the estimation of
timelags becomes difficult if either the signal trace in either KD
time-frequency map is broken into parts (as the signal spans over
several IMFs in the characterization stage), or if noise enters the
KD time-frequency map. A break up of the signal trace over several
IMFs, also known as mode mixing, is a possible outcome if the signal
spans a large dynamic frequency range \cite{huang2005hht}. Event
894398023 shows the interplay of these effects and how it impacts
the KD time-frequency map (Fig. \ref{Flo:Data89439802timelag}). We
find in Fig. \ref{Flo:Data894398023-1} the individual KD time-frequency
maps of this event, with Livingston showing mode mixing. This yields
an timelag estimate with an uncertainty of \textasciitilde{}5 msec
as it is not clear at which point an optimal overlap is achieved,
since part of the trace in Hanford coincides with a gap in Livingston.
Noise affects the time-frequency estimate in two additional ways:
by causing artifact traces in the Hanford and Livingston KD time-frequency
maps, and imposing fluctuations on the signal trace. While this remains
an area of investigation, half of the events show timing uncertainties
less than 1 msec, and the fraction of timelag estimates greater than
4 msec was less than 1/3 of the total; as mentioned earlier.

\section{Outlook and Discussion}

We presented the application of a new data analysis pipeline based
on the Hilbert Huang Transform. Our approach yielded similar sensitivity
to the other pipelines, with a comparable number of detected events
\cite{ninja}. The most significant feature of our pipeline may be
seen in its ability to display the time-frequency evolution of the
event with very high precision, free of the time-frequency uncertainty
of transforms utilizing basis sets (e.g., the Fast Fourier Transform).
These highly resolved KD time-frequency maps open the possibility
to estimate timelags to high accuracy between detectors based on the
maps overlap, and will also allow the possibility of lower detection
threshholds by using a veto based on the overlap in time and frequency
of the time-frequency timelag estimate. Future research will involve
finetuning and further exploration of methods to yield robust and
accurate pipeline results.

\section*{References}

\bibliographystyle{iopart-num}
\bibliography{hhtatlowsnr,ninja}

\end{document}